\documentstyle[epsf,psfig,aps,prl,twocolumn]{revtex}
\begin{document}
\vskip 0.5cm
\title{
How big is the source that produces quark gluon plasma in heavy ion collisions?
}
\author{Ambar Jain and V. Ravishankar}
\vskip 0.5cm
\address{Department of Physics, IIT Kanpur, Kanpur-208 016, INDIA
}
\maketitle
PACS: 12.38.Mh, 25.75.-q,24.85.+p, 25.75.Gz
\begin{abstract}
We study, for the first time, the spatial extension of the "source" that produces
 quark gluon plasma (QGP) in ultra relativistic heavy ion collisions (URHIC). The longitudinal
  dimension is studied as a function of time as the system  evolves. The source size is
  found to exhibit  a novel non-classical feature.
\end{abstract}
\maketitle
\narrowtext
\section{Introduction}
The ultimate aim of studying ultra relativistic heavy ion collisions (URHIC) is
not merely to establish the production of Quark Gluon Plasma (QGP) - by looking at
various signals -  but to get a complete picture of the space-time evolution of QGP. In
short, one hopes to describe the evolution of this unique deconfined state of
hadronic matter in all its stages - production, equilibration
and hadronisation. Ambitious that this  might seem, it has indeed proved impossible
to separate the study of  signals such as $J/\Psi$ suppression or strangeness enhancement
without having an at least approximate idea of the production and evolution of QGP.

We have recently proposed a mechanism \cite{prl03}
 for producing QGP in URHIC -
the soft and the semisoft quarks and gluons that constitute the bulk of the plasma.
The mechanism manifests as the source term in the transport equation
for the distribution functions for the partons. This source term, which  is the rate
at which the partons are produced in the extended phase space,
 has the following features: (i) it is defined in the two particle phase space,
 (ii) the production rate is non-Markovian in time\cite{schmidt},
(iii) the vacuum has a dynamical role to play \cite{bloch} -
it acts both as a source and a sink,
(iv) the partons are quasiparticles with a finite life time \cite{weldon},
(v)  the rate
evaluation takes into account the time scales involved in the production vis-a-vis the
time intervals over which the rate is determined,
(vi) the phase space dependence does  not  violate the uncertainty
principle, and finally, (vii) the dynamical nature of the colour charge
 is manifest. Given all these attributes, it is natural to ask
whether the source term can throw light on the spatial dimensions of the QGP as it evolves in time.
Although a complete answer to this question has to be kept pending until the transport
equation is solved in a self consistent manner, the purport of this paper is to show that
the source term by itself does have  interesting information. We shall illustrate this by looking
at the longitudinal dimensions in a generic example.

One promising observational tool for studying dynamically the spatial extension of
the fireball is the measurement of HBT correlations \cite{hb54,hb56,pur56}
in two and three pions (and kaons)
that are produced in URHIC. In particle physics, this idea was first proposed by
Goldhaber {\it et. al.} \cite{gold60} for $p\bar{p}$ collisions.
 In the context of QGP, a major impetus
for such a study is  the hope that a knowledge of the system size would also
settle whether the source producing the pions/kaons is nuclear, or its
 deconfined state.
A useful tool that it is,  measurements in HBT
 are still beset with many
uncertainties - of extracting and interpreting data. Much of it is due to the fact that
the source here is not static;  it
evolves rapidly,  over time scales of several fermis and over length scales of the
same order. The reconstruction of the source history from HBT observations is,
 therefore,
not straight forward: it has been pointed out  \cite{heinz96,gast00}
 that what gets
determined is not the true "size" of the system, but only the so called regions of
"space time homogeneity" .  The other complications
 are the finite life time of the source, inhomogeneous
temperature profiles and a strong collective dynamical expansion \cite{heinz96}.
At a more basic level, HBT studies have the best utility if  the pions are
produced from a chaotic
source; there is no clinching   justification (theoretically) to
assume that in the case of
URHIC; studies based on the Lund model
\cite{bo86,bia97} indicate that the sources do possess  non-chaoticity.
An empirical test to check the chaoticity is to measure the so
called $\lambda$ parameter in two
pion correlations. However, $\lambda$ itself   depends
on other paramters which are either incompletely or inaccurately known \cite{3pi03}.

Apart from these caveats, the extraction of radii involves additional assumptions:
 the interpretation and parametrisation of HBT data
 assume a Gaussian profile for the source.
 Reasonable and natural that it seems, it still needs justification.
 Next, the (spurious)
  contributions from the Coulomb interaction have to be subtracted.
  More inportantly,
  one needs a criterion for "emission instant" and the  "duration of
  emission of the particles" - both of which
   are ill defined
  in quantum mechanics \cite{prl03,gast00}. The phase space coordinates should
  also be sufficiently
  coarsened so as not to violate uncertainty principle \cite{prl03}.
  These problems are nontrivial to
  handle both in experiments and theoretical simulations \cite{gast00}.
  An illustrative instance of the present situation is  the so called
  RHIC puzzle \cite{starphen12} which is still not understood properly.

  In short, the utility of the HBT analyses gets enhanced if
  they can be supplemented with an independent
  theoretical investigation of the space time dynamics of QGP.
   As mentioned at the beginning, the transport
  studies provide the required  frame work. We shall show that the
   the source term
  proposed in \cite{prl03} does indeed give valuable information
  on the spatial extension of QGP.

  The next section introduces the source term derived in \cite{prl03} and the
  subsequent section discusses the spatial extension of the source. For the
  sake of simplicity, we consider only the longitudinal dimension.

  \section{The source term}
  We refer to \cite{prl03} for the motivation and details of deriving the source term.
  To summarise in a nut shell,
  the production of QGP  takes place via the decay of a mean chromoelectric field (CEF)
  that is produced
  in between the two nuclei soon after they have collided and start receding
  from each other. By
  energy momentum conservation, the CEF acquires a space - time dependence,
  thanks to which the
  instability of the QCD vacuum may be studied perturbatively.
  It may be {\em emphasised} that the existence of CEF is itself
  a consequence of the non-perturbative aspects of QCD. Indeed,
the chromoelectric field may  be considered to be a manifestation  of the
strings in the colour flux tube model \cite{flux}, which is known to provide a natural
setting for discussing quark confinement \cite{thooft}.
The main difference between this work and the conventional approaches
is that the latter do not
take into cognisance the space time dependence of the CEF in invoking
the production mechanism; they
employ Schwinger mechanism \cite{sch} which is valid only if the field is
 uniform and
constant. We make no such assumption here (see also \cite{tdsource} in this context).

Consider the gluons. The lagrangian for pair production may be set up by
expanding the gauge potential
$A_{\mu}^a$ as a sum of the background classical potential $C_{\mu}^a$ and
its fluctuation
$\phi_{\mu}^a$ which is operator valued. Keeping terms quadratic
in the fluctuations in the Yang -Mills
action,  we get
\begin{eqnarray}
L_{2g} &= &-\frac{g}{2} f^{abc}\big [(\partial_{\mu}C_{\nu}^a -
\partial_{\nu}C_{\mu}^a)
            \phi^{\mu b}  \phi^{\nu c} \nonumber \\
       & &  +       (\partial_{\mu}\phi_{\nu}^a -
       \partial_{\nu}\phi_{\mu}^a) (C^{\mu b} \phi^{\nu c}
        +     \phi^{\mu b} C^{\nu c}) \big ]+{\cal O}(g^2).
\end{eqnarray}
Taking into account the Wu Yang ambiguity \cite{wu}, and the studies of
Brown and Weissberger \cite{brown}.
  one can argue that $C_{\mu}^a$ should have an abelian form, at least for non vanishing
  leading order contributions.
  By a suitable gauge choice, we may write
  $C_{\mu}^a = \delta_{\mu, 0} \sum_iC_i(t, \vec{r})\delta_{a,i} $
where the summation runs only over the diagonal generators of the gauge group.
 Note that $C_{\mu}^a$ generates only an electric component.

In determining  the production rate, as  a function of time, the crucial step is
 {\em not} to evaluate the S-Matrix. Instead, we study the time dependent evolution of
the state $ \vert \Psi > (t)$ in the Fock space using the standard Schwinger-Dyson
expansion
for the Unitary operator $U(t,0)$, with the boundary condition
$\vert \Psi> (t=0) = \vert vac >$. The instant $t=0$ is singled out as the moment
of the creation of CEF.
The state   at any time is then projected on to
the two gluon sector.
 The  mass shell condition is imposed as a constraint on the physical states.

Denote the two gluon state as
$\vert gg> \equiv \vert \vec{p}_1,\vec{p}_2; s_1,s_2; c_1,c_2 >$,
labelled by the momentum, spin and colour quantum numbers  respectively.
In the leading order, the production
amplitude may be written as
\begin{eqnarray}
<gg \vert T(t) \vert 0 > = \frac{ig}{(2 \pi)^3} \frac{(E_2 -E_1)}{2 \sqrt{E_1 E_2}}
                          \vec{\epsilon}^{s_1}(\vec{p}_1)
                           \cdot \vec{\epsilon}^{s_2}(\vec{p}_2)f^{a c_1 c_2}\nonumber \\
                         \tilde{C}^{0,a}(E_1 +E_2; \vec{p}_1 + \vec{p}_2; t)
\end{eqnarray}
where $T(t) \equiv U(t,0) -1$. Further,
$$
\tilde{C}_0^a =
              \int^t_0 dt_1 e^{-i(E_1 +E_2)t_1} \int d^3 \vec{r}
              exp (i(\vec{p}_1 + \vec{p}_2)
              \cdot \vec{r})
              C^{0,a}(t_1, \vec{r})
$$
is the incomplete Fourier transform of the gauge field and
$E_i$ are the energies carried by the gluons.
The corresponding expression for the $q\bar{q}$ production is given by
\begin{eqnarray}
<q\bar{q}\vert T(t) \vert 0> = \frac{g}{(2\pi)^3} \frac{m}{\sqrt{E_1E_2}}
                 \tilde{C}_0^a T^a_{c_1, c_2} u^{\dagger}_{s_1}
                 (\vec{p}_1)v_{-s_2}(-\vec{p}_2),
\end{eqnarray}
with  $\vert q\bar{q} > \equiv \vert \vec{p}_1, \vec{p}_2; s_1,s_2; c_1,c_2 >$.
$T^a$ are the generators of the gauge group in the fundamental representation,
while $u, v$ are the usual Dirac
spinors.

The probability that a pair is produced {\it any time}
during the interval $ (0,t)$ is given by
$\vert <\xi \vert T(t) \vert vac > \vert^2$, $\xi$ standing
for either $gg$ or $q\bar{q}$. In
\cite{prl03}, rates were extracted by taking the derivative of the
probability with respect
to time. Here, we study the probabilities directly.

In order to extract the size parameters, we deviate from the evaluation
performed in \cite{prl03}. We label the two parton
states   by their position
coordinates  $\vec{r}_1, \vec{r_2}$,
which entails the standard Fourier traansform of Eqns. 2 and 3 in the
variables $\vec{p_1}, \vec{p_2}$.
 The magnitude squared
of the amplitudes now has the significance that it is the probability that a pair
gets created in the interval $(0,t)$, with the particles found at $\vec{r}_1, \vec{r_2}$
at the instant t. For our purposes we may sum over the spin, colour and one of the
position coordinates. The resulting quantity $P(\vec{r}, t)$ gives the (unnormalised) probability
density for the parton as a function of time. Please note that it is {\it incorrect} to interpret
$t$ as the instant at which the parton is created.

We employ the same CEF that was introduced in \cite{prl03} to study the prdouction rate in
phase space. It is given by
$$
{\cal E}^a_i = \delta_{i,z}{\cal E}_0(\delta_{a,3}+ \delta_{a,8})
 exp\{(\vert z \vert -t ) /t_0 \}
 \theta (t) \theta (t^2-z^2)
 $$
This choice  is close to the boost invariant configuration  required in the Bjorken
scenario \cite{bj}. The gauge group is $SU(3)$, and the field is dependent only on $z,t$.
The probability along the transverse directions is, therefore, uniform, {\it i. e.}, a gaussian
distribution  with a width given given by the diameter of the nucleus. This dimension remains constant
in time.
However, the longitudinal extension
has to be determined explicitly.

\section{The longitudinal extension}
We obtained the production probability density for  gluons and quarks in
the one-particle configuration space by
performing a multi-dimensional fast fourier transform on the momentum space
 amplitude, and integrating over the
unwanted degrees of freedom. The densities are shown in Figs. 1 and 2 as functions
of $z$ for different time instants namely
$t=1,~2,~5,~10,~20$ in units of $t_0$.
The  probability densities are displayed only on the positive
$z$ axis. The negative half is symmetric
about $z=0$. The range of the probabilities (unnormalised!) is  truncated in
order to improve the
visibility of the important features of the curves.
The probability densities  at $z=0$ are large but finite.
These curves shed light on an interesting aspect of the source.

Recall that the CEF has a support in the interval $z \in (-t, +t)$ at any time $t$, with
the two nuclei at $\vert z \vert =t$  moving with unit speed in opposite directions.
We may expect, n\"{a}ively, that the particle production should also be restricted within
the interval $(-t, t)$. The graphs clearly bely these classical expectations.
   The required features are most pronounced in Fig. 1, where
we show the results for the gluons. Strikingly, the probability density
does not terminate
at $z=t$. Instead, it extends beyond, as a broad plateau all the way upto $z=2t$, where
it falls abruptly, and almost discontinuously. This feature is universal for all
the curves,
indicating that the point $z=2t$ is naturally chosen as the boundary by the
system dynamics.
In that sense, we do not have to extract any length scale by curve fittings.

The quark results are shown in Fig. 2, again for times ranging from $t=1$ to $t=20$,
in units of
$t_0$. The results are not that vivid at earlier times. However, the features get more
pronounced as the system evolves so that the plateau structure is clearly
delineated at $t=10 fm$.

The remarkable aspect of these conclusions is that the size of the system
is quintessentially
non-classical. Indeed,
the particles are found in regions where the field identically vanishes.
Since the boundary of the
CEF located at $z= \pm t$ is moving with the speed of light,
causality forbids the particles
produced any time in the interval $(0,t)$ within the range $ z \in (-t,t)$
from reaching the
regions outside the interval. One concludes that the particles
are indeed
produced in the region where CEF vanishes identically!
We suspect that this could be a generic property of the class of CEF that we
have employed. It is
important to note that the plateau position and its extension has
a topological nature: the magnitudes
change smoothly as the sytem evolves with the time, but the location of the
shoulder is pegged at
the value $ \vert z \vert=t$. A straight forward conclusion that one
draws is that longitudinally, the QGP
source is twice as large as the space between the two nuclei.
The fireball extends beyond the two nuclei.

\begin{figure}[htp]
\centerline{\hbox{\psfig{figure=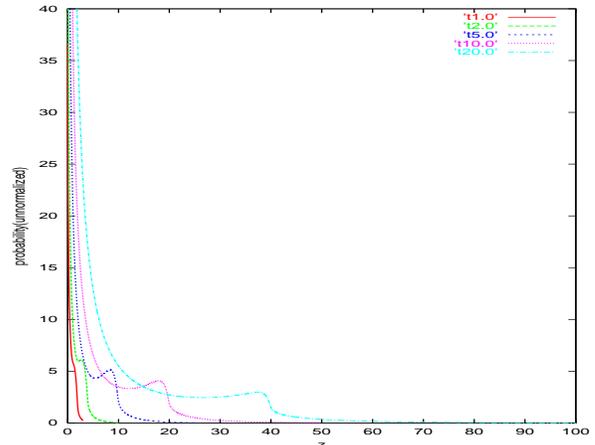,width=8cm,height=6cm}}}
\caption{Unnormalised longitudinal probability density for gluons for times
$t=1,5, 10$ and 20.}
\end{figure}
\begin{figure}[htp]
\vspace*{-0.6cm}
\centerline{\hbox{\psfig{figure=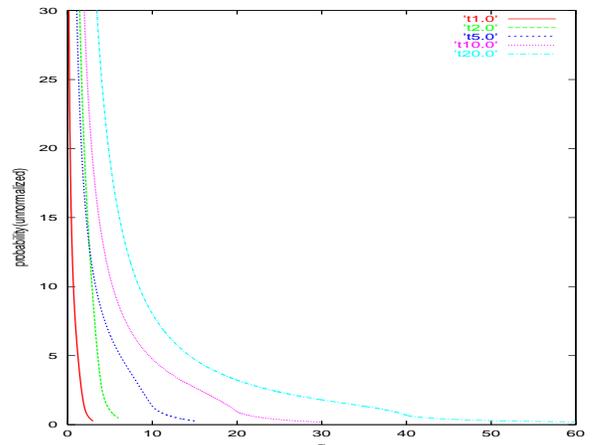,width=8cm,height=6cm}}}
\caption{Unnormalized longitudinal probability density for quarks
as at $t=1,5, 10$ and 20}
\end{figure}

A more quantitative understanding can be obtained by looking at the
relative probability for
finding the particles in the regions $\vert z \vert < t $ and
$ t < \vert z \vert < 2t$. The
ratio for the gluons varies from a maximum of about
$38 \%$ at $t=1.0$, to $15 \%$ at $t=20$.
For the quarks, on the other hand, it varies from
$19 \%$ at $t=1.0$ to $7 \%$ at $t=20$. Clearly,
the non-classical features are the most pronounced when the
field is varying most rapidly, and gets
less important as the field evolves in time. Also, should these
results seem surprising, we may
recall that the probabilities in the regions where the field is
vanishing are, in fact smaller
than the probability that an oscillator in its ground state is
found in the classically forbidden
region. With this perspective, we may conclude that the production
in the regions where the field
vanishes identically is a quantum effect; the occurence of the
step at $ \vert z \vert = 2t$ is
perhaps a feature of the class of fields which have a boost invariant
nature, or are close to it.

A final question remains as to why the spread in the quarks is
less than that of gluons
as indicated by the numbers above.
  The reason for this may be attributed, in part, to the fact that
the two gluon amplitude is antisymmetric in the colour and the
momentum indices of the
two gluons (see Eqn. 2. By Bose symmetry, it is symmetric in the spins).
Consequently, the amplitude will be antisymmetric in the
position variables $\vec{r}_1, \vec{r}_2$ as well.
 The quark case on the other hand, has the charge conjugation symmetry which, when
employed similarly, leads to a symmetric behaviour in the spatial part which
inhibits the spread. In fact, the spread in the momentum probabilities
is larger for quarks than for the gluons, as found in \cite{prl03}.
By the uncertainty
principle, one may expect the quark size to be smaller than that of the gluons.

We thank  Rajesh Gopakumar for a discussion and D. D. Bhaktavatsala Rao
for assistance in preparing the manuscript.

\end{document}